\documentstyle[aps,prl,floats,epsfig]{revtex}

\title{Quantum suppression of shot noise in atom-size metallic contacts}
\author{H.E. van den Brom, and J.M. van Ruitenbeek}
\address{Kamerlingh Onnes Laboratorium, Leiden University, Postbus 9504, 
2300 RA Leiden, The Netherlands}

\begin{document}
\draft

\twocolumn[\hsize\textwidth\columnwidth\hsize\csname@twocolumnfalse\endcsname

\maketitle

\begin{abstract}
The transmission of conductance modes in atom-size gold contacts is investigated 
by simultaneously measuring conductance and shot noise. The results give 
unambiguous evidence that the current in the smallest gold contacts is mostly 
carried by nearly fully transmitted modes. In particular, for a single-atom 
contact the contribution of additional modes is only a few percent. In contrast, 
the trivalent metal aluminum does not show this property.
\end{abstract}

\pacs{PACS numbers: 72.70.+m, 72.15.Eb, 73.23.Ad, 73.40.Jn}

\vskip2pc]

\narrowtext

\newcommand{\av}[1]{\mbox{$\langle #1 \rangle$}}

In 1918 Schottky mentioned shot noise as a fundamental shortcoming of vacuum 
diodes. He realized that the discreteness of electron charge, $e$, causes the 
current to be a Poisson process, and calculated the corresponding mean square 
current fluctuations to be equal to the product of $e$ and the average current, 
$I$, divided by the total time of averaging \cite{schottky}. This type of noise 
is present in all kinds of devices, including microscopic conductors. In the 
last decade it has become clear that it can actually be used to obtain 
information on the electron transport mechanism 
\cite{israel,saclay,lesovik,butt90,bh,ml,butt92,bbjn,utrecht,smd,frac}. For 
example, in a ballistic quantum point contact (QPC) in a two-dimensional 
electron gas (2DEG), the conductance, $G$, as a function of contact diameter 
shows a step-wise increase by integer multiples of the conductance quantum, $G_0 
\equiv 2e^2/h$ \cite{ww}. Recent 2DEG experiments showed that shot noise was 
strongly suppressed at quantized conductance values \cite{israel,saclay}, in 
accordance with theoretical predictions \cite{lesovik,butt90,bh,ml,butt92}. In 
this Letter, measurements of shot noise are performed for the first time to 
analyze the electronic transport properties of atom-size metallic contacts. 

For a metal, the size of an atom is comparable to half the Fermi wavelength 
$\lambda_F$ of the conduction electrons. Therefore, the equivalence of 
electronic properties of QPCs in a 2DEG and in a metal is far from trivial.
In particular, it inhibits a direct observation of the effect of the formation 
of discrete electron modes in a metallic QPC, i.e.\ 
quantization of the conductance. In fact, using a combined scanning tunneling 
microscopy (STM) and atomic force microscopy (AFM) setup \cite{rav} it has been 
shown that steps in the conductance, observed when stretching the contact, are 
the result of atomic rearrangements (see also \cite{scan}). Primarily, evidence 
for quantization of the conductance in metals is derived from histograms of the 
conductance values, which, for gold \cite{br} and sodium \cite{nature}, show 
peaks close to integer multiples of $G_0$. However, this evidence is not 
unambiguous, as demonstrated for aluminum, which shows clear peaks near 
quantized conductance values in the histograms \cite{alex}, while up to three 
modes contribute to the conductance near $G=G_0$ \cite{andreev}.

In a ballistic QPC with perfect transmission of electrons there are no 
fluctuations in the occupation numbers of left and right moving electrons, 
suppressing all shot noise \cite{lesovik,butt90,bh,ml,butt92}. For a contact of 
size comparable to the Fermi wavelength, $\lambda_F$, electron transport is 
described using the 
Landauer-B\"uttiker formalism. In this formalism the shot noise spectral density 
can be expressed in terms of the transmission probabilities $T_n$ of the 
conducting channels \cite{butt90},

\begin{equation}
P_I = 2eV  \frac{2e^2}{h} \sum_n T_n (1-T_n)  ~,
\label{eq:sn} \end{equation}

\noindent where the sum is over all modes which fulfill the quantization 
condition set by the boundaries of the point contact. The label $I$ indicates 
that we consider current fluctuations \av{\Delta I^2}. From Eq.\,(\ref{eq:sn}) 
we 
see that for a mode with a transmission probability, $T_n$, close to 1 the shot 
noise is indeed suppressed. If all $T_n$ are small, the classical full shot 
noise formula, $P_I=2eI$, is recovered. Including the thermal excitations of 
quasiparticles for finite temperatures leads to the following expression 
\cite{ml,butt92},

\begin{eqnarray}
P_I = 2eV \coth \left( \frac{eV}{2kT} \right) \frac{2e^2}{h} \sum_n T_n (1-T_n) 
\nonumber \\
+ ~ 4kT ~ \frac{2e^2}{h}  \sum_n T_n ^2 ~.
\label{eq:coth} \end{eqnarray}

\noindent In equilibrium (i.e.\ $V=0$) or in the high temperature limit, the 
Johnson-Nyquist thermal noise, 4kTG, is recovered. For zero temperature, 
eq.(\ref{eq:coth}) reduces to the shot noise formula (\ref{eq:sn}). In the 
experiments described below we are interested in the amount of noise above the 
thermal noise, which we will call excess noise, $P^{\,\text{exc}}(I) \equiv 
P(I,T) - P(0,T)$.

In order to obtain a stable atomic scale contact, we use the mechanically 
controllable 
break-junction technique (MCB). A notched, 99.998\% pure gold wire is glued on 
top of a phosphor bronze substrate, which is insulated with kapton foil. This is 
mounted into a vacuum can and cooled down to 4.2\,K. By bending the substrate 
the wire is broken, after which contact between the fracture surfaces is 
controlled using a 
piezo-electric element. For a more complete description of the technique see 
Ref.\,\cite{chr}.

\begin{figure}[!t]
\begin{center}
\leavevmode
\epsfig{figure=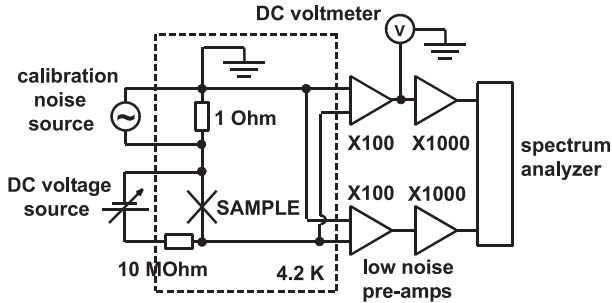,width=8cm,angle=0}
\end{center}
\caption{Wiring diagram of the experimental setup for shot noise measurements on 
atom-size point contacts. The area enclosed by the dashed lines indicates the 
part at 4.2\,K.}
\label{fig:circuit}
\end{figure}

The noise level we are expecting is on the order of nV/$\sqrt{\text{Hz}}$. In 
order to measure such small signals we use the experimental setup schematically 
drawn in Fig.\,\ref{fig:circuit}. The signal to be measured is first amplified 
$10^5$ times by two stages of low noise wide band 
pre-amplifiers (DC $-$ 1\,MHz and 0.5\,Hz $-$ 1\,MHz, respectively). In order to 
suppress the noise of the 
pre-amplifiers we use two sets of 
pre-amplifiers in parallel and feed the signals into a spectrum analyzer, which 
calculates the cross-correlation Fourier spectrum. The conductance of the sample 
is measured using the DC voltage after the first amplifier; the bias current is 
sent using a battery (0$-$9\,V) with a large series resistance, the latter being 
at 4.2\,K close to the sample. Since we are interested in contacts with 
resistance around 13\,k$\Omega$, and our wiring has a capacitance of around 
250\,pF, this introduces a low pass filter with a 
cut-off frequency of around 50\,kHz. To correct the measured spectra, $P_V(I)$, 
for the electronic transfer function of the system, we send white noise, 
$P_V^{\,\text{src}}$, from a calibration source through a 1\,$\Omega$ series 
resistance and measure the frequency response of the setup $P_V^{\,\text{cal}}$. 
The excess noise then equals

\begin{equation}
P_V^{\,\text{exc}}(I) =  \frac{P_V(I) - P_V(0)} 
{P_V^{\,\text{cal}} - P_V(0)} 
\times P_V^{\,\text{src}}
~.
\label{eq:corr} \end{equation}

\noindent The label $V$ indicates that here we consider the spectral density of 
voltage fluctuations \av{\Delta V^2} \cite{PIPV}. 

Obtaining one single spectrum in the interval 250\,Hz $-$ 100\,kHz and averaging 
10000 times takes about 1 minute. For each contact setting, we have to take 
several spectra: one for calibration, one for thermal noise, and several with 
different DC current biases. During this procedure, the contact is verified to 
be stable by measuring the conductance between each two spectra. Measuring the 
conductance twice with different polarity of the bias current eliminates DC 
offset of the 
pre-amplifiers.

For frequencies smaller than the inverse dephasing time and obeying $hf<<eV,kT$ 
the spectrum of both thermal and shot noise is white, i.e.\ frequency 
independent. This is not the case for 
so-called 1/f noise, which is generally believed to be a consequence of defect 
motion and decreases at higher frequencies as $f^{-1}$. External vibrations 
(e.g.\ sound) and 
pick-up of external electromagnetic signals can be recognized in the Fourier 
spectrum. External vibrations are mostly visible at low frequencies, in the same 
range where 1/f noise dominates, which is why we concentrate on the higher 
frequency part of the spectrum. Electromagnetic 
pick-up, present despite careful shielding, is seen as sharp peaks and is 
removed when we take the difference (\ref{eq:corr}). In addition, the 
pre-amplifier noise left over after taking the cross correlation is further 
removed this way. Furthermore, the effect of external vibrations as well as 1/f 
noise is quadratic in the supplied current, while shot noise is linear. We 
verify that the excess noise we measure has the proper current dependence.

\begin{figure}[!b]
\begin{center}
\leavevmode
\epsfig{figure=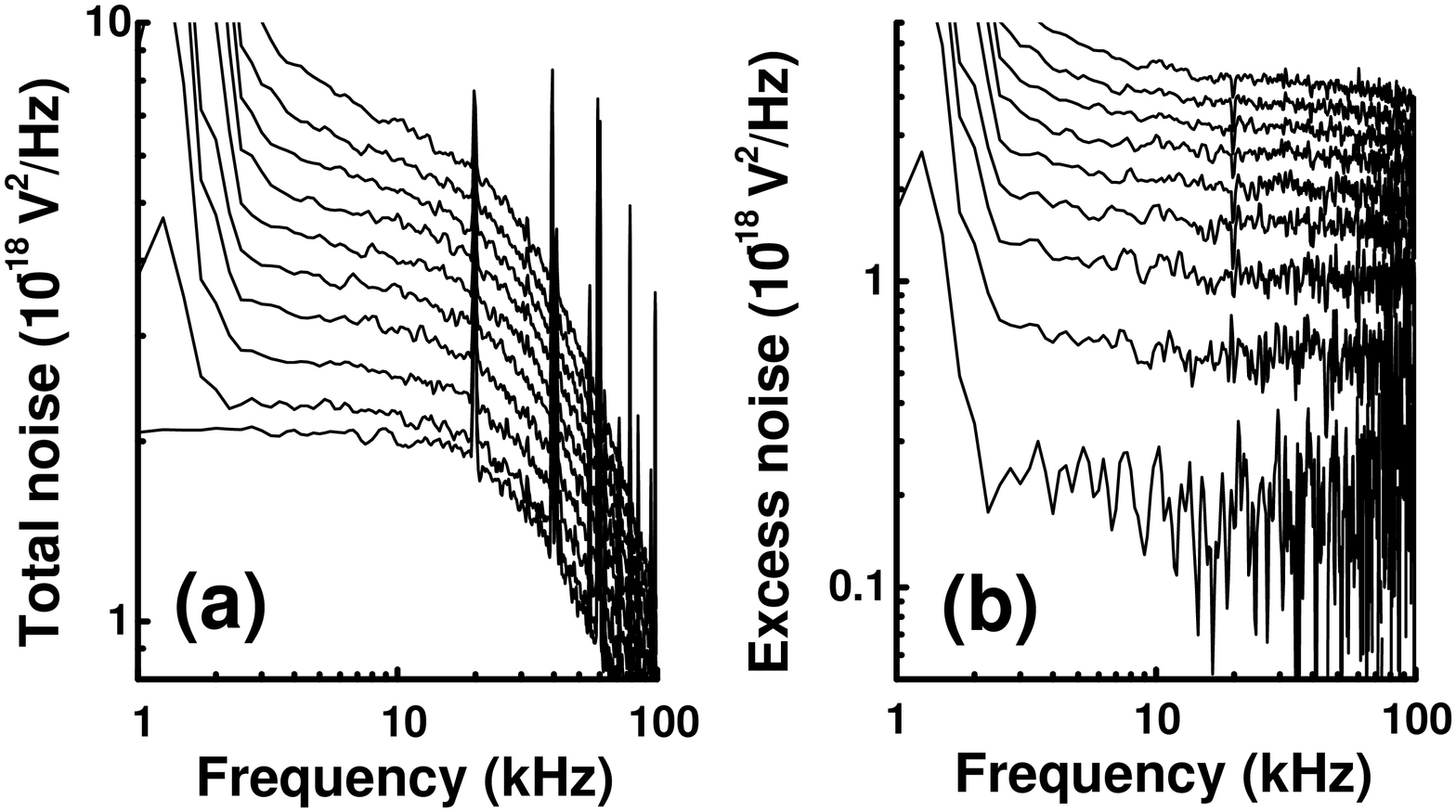,height=8cm,angle=270}
\end{center}
\caption{a) Raw data for a gold contact of 1.53\,$G_0$ at 4.2\,K, without 
correction for the electronic transfer of the set-up. The lowest curve shows the 
thermal noise, a few percent higher than the theoretical value 
2.0\,\,$10^{-18}\,\text{V}^2/$Hz, which is due to the preamplifier noise. The 
upper curves are the total measured noise for increasing current, $I=0.1, 0.2, 
0.3, \dots 0.9\,\mu$A. b) Excess noise, calculated from the data in 
Fig.\,\ref{fig:rawexcess}a and corrected according to Eq.\,(\ref{eq:corr}).
Note that the effect of mechanical vibrations is visible at lower frequencies.} 
\label{fig:rawexcess}
\end{figure}

\begin{figure}[!t]
\begin{center}
\leavevmode
\epsfig{figure=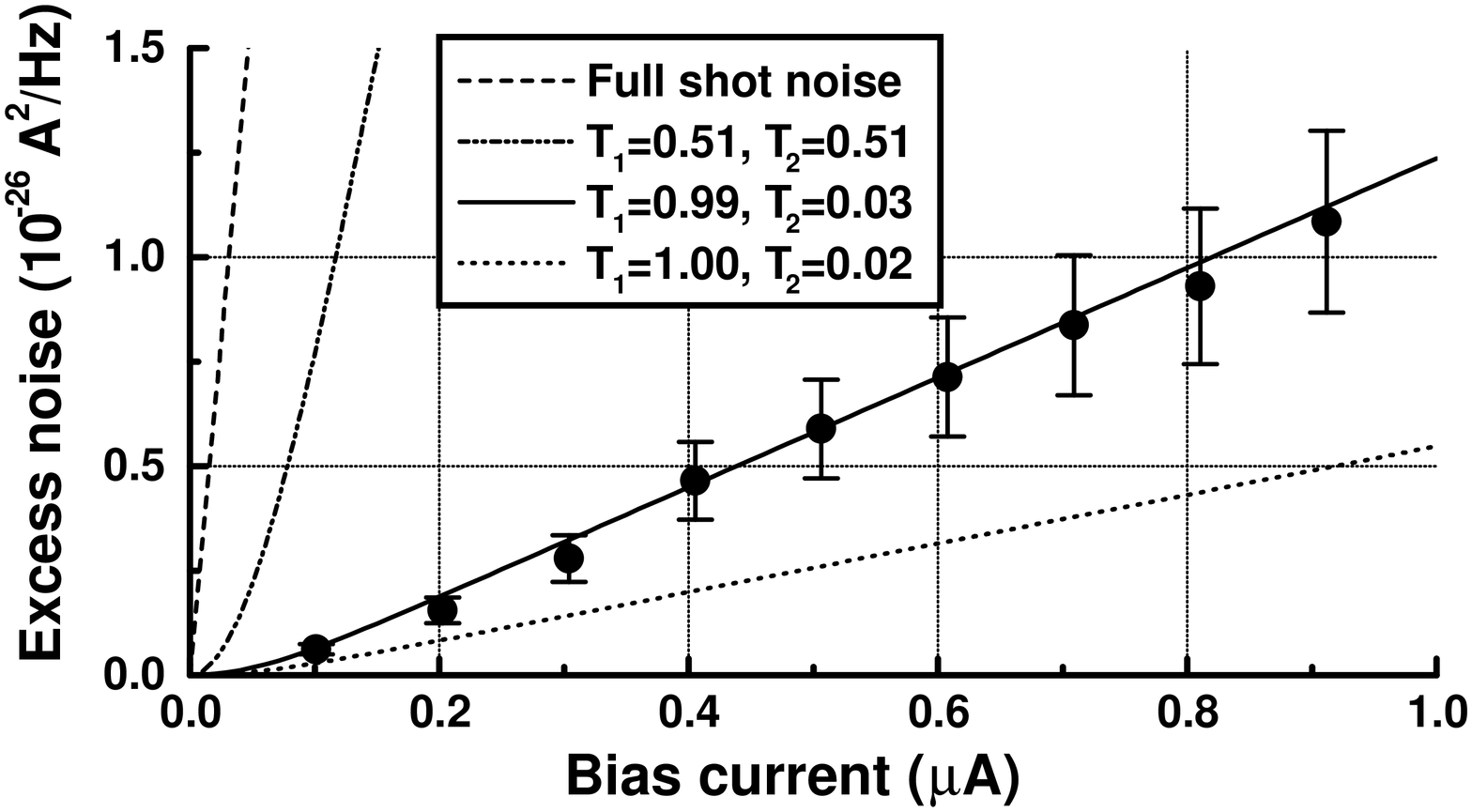,height=8cm,angle=270}
\end{center}
\caption{Measured excess noise values for a contact with $G=1.02\,G_0$, as a 
function of bias current. For comparison we plot full shot noise, $2eI$, and 
Eq.(\ref{eq:coth}) for the case of one single partially transmitted wave (i.e.\ 
$T_1=1$ and $T_2=0.02$) and for the case of two equally transmitted waves 
($T_1=T_2=0.51$). A good description of the data is obtained with $T_1=0.99$ and 
$T_2=0.03$.}
\label{fig:CompP(I)}
\end{figure}

\begin{figure}[!b]
\begin{center}
\leavevmode
\epsfig{figure=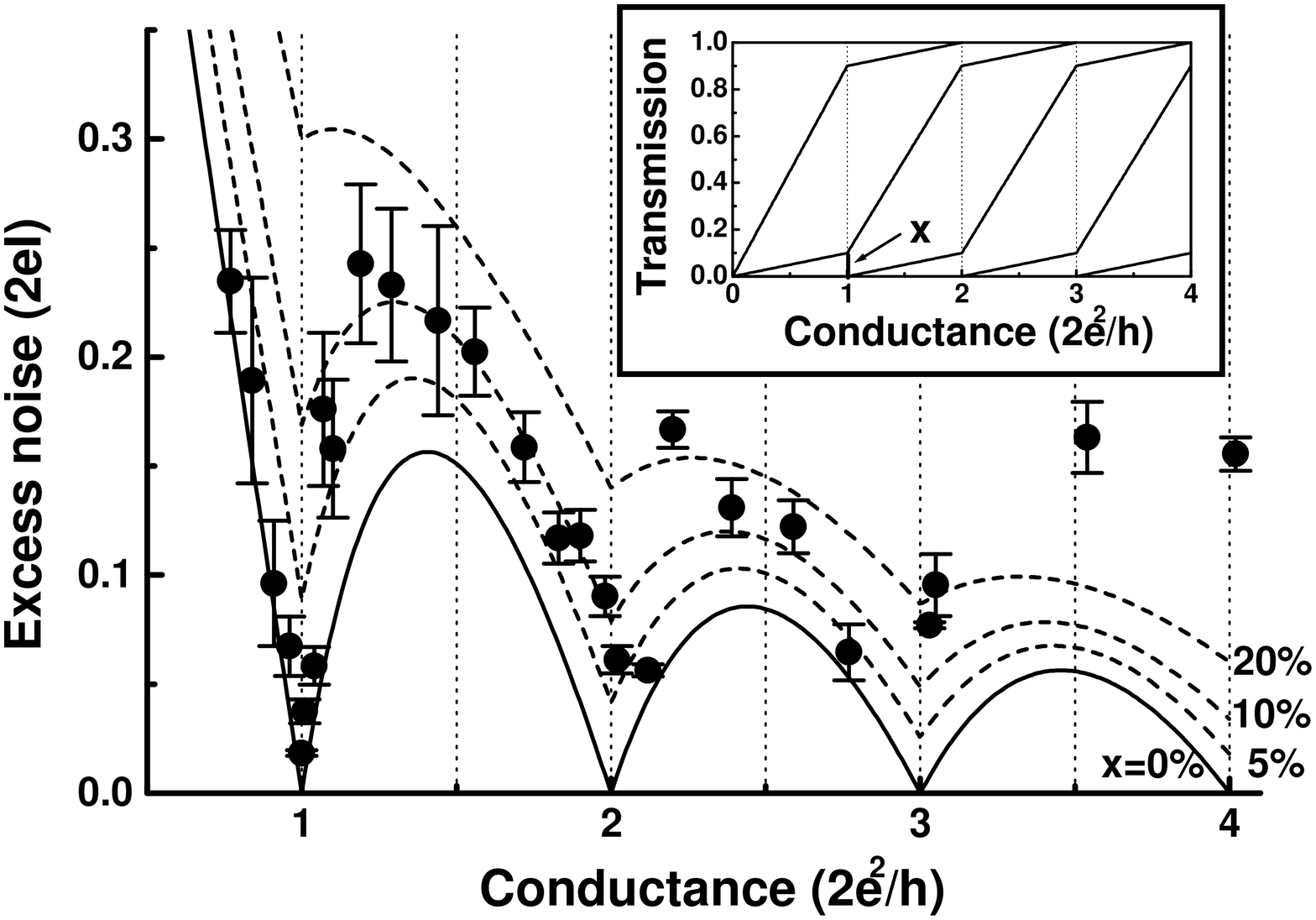,height=8cm,angle=270}
\end{center}
\caption{Measured excess noise values for 27 gold contacts at 4.2\,K with a bias 
current of 0.9\,$\mu$A. Comparison is made with calculations in the case of one 
single partially transmitted mode (full curve) and for various amounts of 
contributions of other modes according to the model described in the inset 
(dashed curves). In the limit of zero conductance, these curves all converge to 
full shot noise, i.e.\ 2.9\,\,10$^{-25}\,$A$^2$/Hz. {\it Inset:} transmission of 
modes in the case of x=10\% contribution from neighboring modes.}
\label{fig:P(G)}
\end{figure}

A typical example of the raw data, $P_V(I)$, we obtain can be seen in 
Fig.\,\ref{fig:rawexcess}a, where we show measurements on a contact of 
8.4\,k$\Omega$, or 1.53\,$G_0$. The zero bias noise at low frequencies is in 
agreement with the expected thermal noise, $4kTR$. At higher frequencies, we 
observe a 
roll-off, which is due to the frequency response of the setup. When applying a 
bias current the noise level increases. Apart from an increase in noise level 
over the entire spectrum, we see a rapid rise at low frequencies, which we 
attribute to 1/f noise and external vibrations. The excess noise, 
$P_V^{\,\text{exc}}(I)$, corrected for the transfer function according to 
Eq.(\ref{eq:corr}), is plotted in Fig.\,\ref{fig:rawexcess}b and approaches a 
white spectrum for higher frequencies. At the highest frequencies, a small 
decrease is observed due to higher order stray capacitance corrections, which we 
take into account in the error bars of Fig.\,\ref{fig:CompP(I)} and 
\ref{fig:P(G)}. To show that we are indeed measuring shot noise we investigated 
the current dependence and verified the expected behavior given by 
Eq.(\ref{eq:coth}), which is nearly linear for higher current: In 
Fig.\,\ref{fig:CompP(I)} we compare the data for a contact with $G = 1.02\,G_0$ 
to 
full shot noise, $2eI$, and to Eq.\,(\ref{eq:coth}) with different combinations 
of 
transmission values. As can be seen in the figure, the data are well described 
by one almost fully transmitted mode together with one mode with a very small 
transmission probability.

A number of contacts with conductances in the range $0.7 - 4.1\,G_0$ have been 
similarly analyzed. In Fig.\,\ref{fig:P(G)} we plot the excess noise values 
corresponding to the measured voltage noise at maximum bias current ($0.9\,\mu 
A$) as a function of conductance for 27 different contacts. The figure shows 
that all values are small, compared to $2eI$. A smallest value of $0.02 \pm 
0.005\,(2eI)$ is observed for a contact with conductance very near $G_0$. For 
comparison we show the expected behavior, based on Eq.\,(\ref{eq:coth}), when 
the conductance is due to only fully transmitted modes ($T_n=1$) plus a single 
partially transmitted mode (full curve). All measured points are at or above 
this curve, as is expected. From our measurements we cannot determine the 
complete set $\{T_n\}$, since we have only two measured parameters (i.e.\ 
conductance and shot noise) and we can obtain at most two independent parameters
$T_n$. To visualize the effect of contributions of different modes to the 
conductance we use the model described in the inset of Fig.\,\ref{fig:P(G)}. In 
this model the conductance between $(n-1)\,G_0$ and $n\,G_0$ is built up as $G = 
(n-2)\,G_0 + (T_{n-1} + T_n + T_{n+1})\,G_0$, where the three partially open 
channels have transmissions which increase linearly, and the sum of $(1-T_{n-
1})$ and $T_{n+1}$ is a constant fraction $x$. This model has no physical basis 
but serves merely to illustrate the extent to which additional, partially open 
channels are required to describe the measured shot noise. The corresponding 
behavior of the excess noise as a function of conductance, calculated from 
Eq.\,(\ref{eq:coth}), is shown as the dashed curves in Fig.\,\ref{fig:P(G)} for 
$x=5\%, 10\%$ and $20\%$. We see that for $G<G_0$ the data are very close to the 
$x=0\%$ curve, while for $G_0<G<2\,G_0$ the data are closer to the $x=10\%$ 
curve. In particular the minimum at $G=G_0$ is very sharp, while also a minimum 
near $G=2\,G_0$ is visible. For $G>2\,G_0$ the contribution of other
partially open channels 
continues to grow. Each point is measured for a different contact, and the 
contribution of modes to the conductance of each contact can in principle be 
widely different. For example, just above $2\,G_0$ we find a point with $x<5\%$, 
while the next point has $x>20\%$. Hence the scatter in the data compared to any 
of the curves is not due to statistical errors but is a result of the intrinsic 
variation in the properties of the contact. 

It should be stressed that the results described above are obtained in a gold 
QPC. We performed similar experiments on aluminum, which showed much weaker 
suppression of shot noise. For Al contacts between 0.8\,$G_0$ and 2.5\,$G_0$ the 
obtained shot noise values vary from 0.3 to 0.6\,($2eI$). These values only fit 
the theory if one assumes contributions from a much wider set of modes. In 
particular, the conductance at $G=G_0$ is not carried by one fully transmitted 
mode but by at least two partially transmitted modes instead.

In model simulations for gold Brandbyge {\it et al.} \cite{bsj} find nearly full 
transmission of the channels for $G$ near 1 and 3\,$G_0$. However, they report 
two half opened channels around $G=2\,G_0$, which is at variance with the 
present experimental results. 

For atomic-size contacts of superconductors, Scheer {\it et al.} 
\cite{andreev,chem} performed experiments on current-voltage characteristics in 
the superconducting state. This was used to demonstrate that the number of 
conducting channels in a single atom contact is determined by the number of 
valence orbitals \cite{chem,clm}. However, reservations were made about the 
results on gold, since the proximity effect was used to induce superconductivity 
in a gold QPC, causing a modified quasiparticle density of states. Our results 
for $G<G_0$ show unambiguously that in the monovalent metal gold, the current 
through a single atom contact is indeed almost exclusively carried by one single 
conductance channel. Our results for a single atom contact of aluminum (which 
has a valency of three) confirm that several modes are transmitted 
\cite{andreev,chem,clm}. 
In addition, for gold we find that the conductance for larger contacts is 
well-described by a set of conductance channels, where all are fully open, 
except one, which carries the fraction of conductance above the integer value. 
This property was also inferred from measurements of conductance fluctuations 
\cite{2nd} and thermopower \cite{thermo} by Ludoph {\it et al.}, and was 
referred to as ``saturation of channel transmission''. From Fig.\,\ref{fig:P(G)} 
we can even quantify the accuracy to which this rule is obeyed. Where the 
interpretation of Refs. \cite{2nd,thermo} depends on ensemble averaged 
properties, the present results show that saturation of channel transmission is 
observed for {\it individual} contacts, and is independent of any adjustable 
parameters. 

From conductance histograms the quantum conductance properties cannot be 
identified as unambiguously in metals \cite{alex} as they are in 2DEG contacts. 
In gold QPCs the conductance does not show the pronounced preference for integer 
multiples of $G_0$ \cite{br,gai} as would be expected for truly quantized 
conductance. In contrast, the quantum suppression of shot noise observed here, 
is even more pronounced than in 2DEG systems \cite{israel,saclay}. By this 
method the quantum nature of the conductance in atom-size metallic point 
contacts is now straightforwardly revealed.

This work is part of the research program of the Stichting voor Fundamenteel 
Onderzoek der Materie (FOM), which is financially supported by NWO. We thank 
D.C. Glattli, J.I. Dijkhuis and M.G. Peters for helpful discussions and L.J. de 
Jongh for his stimulating interest in this work.



\begin{thebibliography}{99}
\bibitem{schottky} W. Schottky, Ann.\ Phys.\ (Leipzig) {\bf 57}, 541 (1918)
\bibitem{israel} M. Reznikov {\it et al.}, Phys.\ 
Rev.\ Lett.\ {\bf 75}, 3340 (1995)
\bibitem{saclay} A. Kumar {\it et al.}, Phys.\ Rev.\ Lett.\ {\bf 76}, 2778 
(1996)
\bibitem{lesovik} G.B. Lesovik, Sov.\ Phys.\ JETP Lett.\ {\bf 49}, 592 (1989)
\bibitem{butt90} M. B\"uttiker, Phys.\ Rev.\ Lett.\ {\bf 65}, 2901 (1990)
\bibitem{bh} C.W.J. Beenakker and H. Van Houten, Phys.\ Rev.\ B {\bf 43}, 12066 
(1991)
\bibitem{ml} Th. Martin and R. Landauer, Phys.\ Rev.\ B {\bf 45}, 1742 (1992)
\bibitem{butt92} M. B\"uttiker, Phys.\ Rev.\ B {\bf 46}, 12485 (1992)
\bibitem{bbjn} C.W.J. Beenakker and M. B\"uttiker, Phys.\ Rev.\ B {\bf 46}, 1889 
(1992); M.J.M. de Jong and C.W.J. Beenakker, Phys.\ Rev.\ B {\bf 46}, 13400 
(1992); K.E. Nagaev, Phys.\ Lett. A {\bf 169}, 103 (1992)
\bibitem{utrecht} F. Liefrink {\it et al.}, Phys.\ Rev.\ B {\bf 49}, 14066 
(1994)
\bibitem{smd} A.H. Steinbach, J.M. Martinis, and M.H. Devoret, Phys.\ Rev.\ 
Lett.\ {\bf 76}, 3806 (1996)
\bibitem{frac} L. Saminadayar {\it et al.}, Phys.\ Rev.\ Lett.\ {\bf 79}, 2526 
(1997)
\bibitem{ww} B.J. van Wees {\it et al.}, Phys.\ Rev.\ Lett.\ {\bf 60}, 848 
(1988); D.A. Wharam {\it et al.}, J. Phys.\ C {\bf 21}, L209 (1988)
\bibitem{rav} G. Rubio, N. Agra\"\i t, and S. Vieira, Phys.\ Rev.\ Lett.\ {\bf 
76}, 2302 (1996)
\bibitem{scan} H.E. van den Brom, A.I. Yanson, and J.M. van Ruitenbeek, Physica 
B {\bf 252}, 69 (1998)
\bibitem{br} M. Brandbyge {\it et al.}, Phys.\ Rev.\ B {\bf 52}, 8499 (1995)
\bibitem{nature} J.M. Krans {\it et al.}, Nature {\bf 375}, 767 (1995)
\bibitem{alex} A.I. Yanson, and J.M. van Ruitenbeek, Phys.\ Rev.\ Lett.\ {\bf 
79}, 2157 (1997)
\bibitem{andreev} E. Scheer {\it et al.}, Phys.\ Rev.\ Lett.\ {\bf 78}, 3535 
(1997)
\bibitem{chr} C.J. Muller, J.M. van Ruitenbeek, and L.J. de Jongh, Physica C 
{\bf 191}, 485 (1992); Phys.\ Rev.\ Lett.\ {\bf 69}, 140 (1992)
\bibitem{PIPV} The relation between current and voltage noise is simply $P_V = 
P_I/G^2$.
\bibitem{bsj} M. Brandbyge, M.R. S\char'34rensen, and K.W. Jacobsen, Phys.\ 
Rev.\ B {\bf 56}, 14956 (1997)
\bibitem{chem} E. Scheer {\it et al.}, Nature {\bf 394}, 154 (1998)
\bibitem{clm} J.C. Cuevas, A. Levy Yeyati and A. Mart\'\i n-Rodero, Phys.\ Rev.\ 
Lett.\ {\bf 80}, 1066 (1998) 
\bibitem{2nd} B. Ludoph, D. Esteve, C. Urbina, M.H. Devoret, and J.M. van 
Ruitenbeek, Phys. Rev. Lett. (this issue) 
\bibitem{thermo} B. Ludoph and J.M. van Ruitenbeek, Phys. Rev. B, submitted.
\bibitem{gai} Z. Gai {\it et al.}, Phys.\ Rev.\ B {\bf 53}, 1042 (1996)
\end{thebibliography}
\end{document}